# Comparison Analysis of Tree Based and Ensembled Regression Algorithms for Traffic Accident Severity Prediction

## Muhammad Umer, Saima Sadiq, Abid Ishaq, and Saleem Ullah
Department of Computer Science, Khwaja Fareed University of Engineering and Information Technology, Rahim Yar Khan, 64200, Pakistan, umersabir1996@gmail.com

## Najia Saher
Department of CS & IT, Islamia University Bahawalpur, Bahawalpur, Pakistan., najiasaher@gmail.com

## Hamza Ahmad Madni
Department of Computer Engineering, Khwaja Fareed University of Engineering and Information Technology, Rahim Yar Khan, 64200, Pakistan., 101101770@seu.edu.cn

Rapid increase of traffic volume on urban roads over time has changed the traffic scenario globally. It has also increased the ratio of road accidents that can be severe and fatal in the worst case. To improve traffic safety and its management on urban roads, there is a need for prediction of severity level of accidents. Various machine learning models are being used for accident prediction. In this study, tree based ensemble models (Random Forest, AdaBoost, Extra Tree, and Gradient Boosting) and ensemble of two statistical models (Logistic Regression Stochastic Gradient Descent) as voting classifiers are compared for prediction of road accident severity. Significant features that are strongly correlated with the accident severity are identified by Random Forest. Analysis proved Random Forest as the best performing model with highest classification results with 0.974 accuracy, 0.954 precision, 0.930 recall and 0.942 F-score using 20 most significant features as compared to other techniques classification of road accidents severity.

*Key words*: Road accidents severity, Random Forest, Feature Importance Ensemble Learning *History*:

# 1. Introduction

Road traffic accidents are a major cause of injuries, deaths, permanent disabilities and property loss. It not only affects the economy it also affects the health care system because it puts a burden on the hospitals. Statistics shown by the ministry of public security of china from year 2009 and 2011, traffic accidents caused an average of 65123 people to lose their life and 255540 got injuries annually WHO (2018). Identification of primary factors affecting road accident severity is required to minimize level of accidental severity. Accidental Severity does not happen by chance; there are patterns that can be predicted and prevented. Accidental events can be analyzed and avoided Gissane (1965). Being one of the major issues of accident management, accident severity prediction plays an important role to the rescuers to evaluate the level of severity in traffic accidents, their potential impact and implement the efficient accident management procedures.

In the last two decades, accidental severity is one of the popular research areas. Researchers were applying different statistical approaches for road accident classification. These techniques help in analyzing the cause of the road accidents. Mixed logit modeling approach Haleem, Alluri, and Gan (2015), logit model Bedard et al. (2002) and ordered Probit model Zajac and Ivan (2003), are few of traditional statistical-based studies. But these approaches lack the capability to handle multidimensional datasets Chen and Jovanis (2000). At present, due to the large number of available datasets machine learning surpasses the traditional statistical approaches in prediction Sarkar et al. (2019). In the recent past, many researchers have focused on the work related to severity prediction of traffic accidents. The focus of the many researchers' work is to find out the main factors that have an impact on the severity of the traffic accidents. Non parametric models, linear models like data mining techniques have been widely utilized to conduct such analysis. To describe the literature about severity prediction of the traffic accidents researches, their techniques and adopted methods have been discussed.

# 2. Literature Review

Data mining is immensely used in different fields like deep learning has been utilized by many researchers for image classification Umer et al. (2020), text mining Umer et al. (2020), Fake news detection Umer et al. (2020a), and text classification Sadiq et al. (2020), Imtiaz et al. (2020). Traffic accident data analysis using different data mining approaches has been considered by the many researchers. Many works in the literature explored road accident severity in different countries Sameen and Pradhan (2017), Seid et al. (2019), Singh (2017). Authors utilized the ANN to model injury severity of traffic accidents using classifying the injury severity into five different categories (no injury, possible injury, minor non-incapacitating injury, incapacitating and fatality). There were 150 parameters out of which they have selected the most significant one I, e, 16 parameters using some parameter selection algorithms that affect the injury level of the drivers. ANN was deployed to classify the injury severity level. They have achieved the accuracy of 40.71% which is quite low Delen, Sharda, and Bessonov (2006).

For any kind of data analysis regression models are the basic components with the relationship between explanatory variable and response variable. In Kononen, Flannagan, and Wang (2011), authors worked on the severity of traffic accidents in the United States using logistic regression. At probability cut point 0.20 they obtained the model sensitivity and specificity were 40% and 98% respectively. Their research finding also shows that velocity, seat belt use, and crash direction are important parameters for the prediction of severity of traffic accidents.

A classification and regression models tree (CART) is an important data mining technique. It is a non parametric technique. Many researchers used CART as a tool for classification and for prediction in research related to traffic accidents. Kashani and Mohaymany (2011) utilized CART to analyze the rural traffic accidents in Iran. CART is utilized to find the most important variable that affects the severity of the accidents. In the analysis process three classes were predicted in the form of a group of binary prediction models that assists to achieve the higher accuracy for the predicted model, and obtained accuracy is 60.94%.and the important variable which affect the severity of the accidents is improper overtaking and not fasting the seat belt.

Sharma et al. Sharma, Katiyar, and Kumar (2016) used a Support Vector Machine and multi layer perception to analyze road accidents. They experimented on a limited number of data samples. They explored only two variables that are speed and alcohol as a key contributor in road accidents. SVM outperformed with 94% accuracy. They claimed high speed driving after drinking is the reason for the incident. Tiwari et al. Tiwari, Kumar, and Kalitin (2017) used machine learning models like Decision Tree (DT), Naive Bayes(NB) and Support Vector Machine (SVM) for classification and SOM and K-modes for clustering. They achieved better results with cluster dataset.

AlMamlook et al. AlMamlook et al. (2019) used NB, AdaBoost, Random Forest (RF) and Logistic Regression (LR) to find highways with high risk of accident for traffic agencies. They evaluated their models using AUC,ROC, Recall, Precision and F- measure. RF outperformed with 75% accuracy. In another work, Beshah et al. Beshah and Hill (2010) experimented to analyze important road way related variables that can affect road accident severity. They used DT, NB and KNN to make decision rules for road safety measures. They mainly focus on drivers and pedestrians without giving importance to any other factor such as whether, time or speed. They even did not focus on the influence of machine learning model accuracy for better identification of accidental severity risk.

However, severity prediction of road accidents is still under development. In the past work we have seen the room to improve classification accuracy using machine learning models for road safety. There is a dearth of comparing state of the art machine learning models with hybrid models. Obtaining an appropriate approach will improve prediction accuracy. Finding the best paradigm also helps in identifying factors affecting road accidents. Furthermore factors that are more specific to the target can help machine learning models to improve in prediction results that were not identified previously. The aim of the researches related to the traffic accident data govern on data mining can be divided into two categories:

- Prediction of traffic accidents severity
- The important factors that affect the severity of accident

This paper makes use of ensemble learning models to accurately predict the severity of road accidents. The ensemble learning models used in this experiment are AdaBoost Classifier(AB), Gradient Boosting Machine(GBM), Random Forest(RF), Extra Tree(ET), and Voting Classifier(VC). The voting classifier used in this experiment is an ensemble of two machine learning regression models (Logistic Regression and Stochastic Gradient Descent). The proposed model is applied to US road accident dataset in two phases. In the first phase, all 48 features of the dataset are used to predict the severity of the accident. In the first experimental phase, we also calculate the feature importance value of all features using the random forest classifier. In the second phase of the experiment, the top 20 features that are calculated using random forest classifier feature importance are used to train all machine learning models and to predict the severity of road accidents. This research serves the following key contributions:

- Comparative analysis of Tree based and regression based Ensemble learning classifiers such as: AdaBoost Classifier(AB), Gradient Boosting Machine(GBM), Random Forest(RF), Extra Tree(ET), and Voting Classifier(VC).
- Influence of significant variable on evaluation measures such as accuracy, precision and recall is analyzed.
- Most suitable method and suitable input parameters are explored for classification of road accidents' severity.
- As decrease of input variables improves the performance, it will also reduce the cost of data collection.

The rest of the paper is organized as follows. Section 3 presents an overview of the methodology adopted for the current research as well as as detailed description of the US road accident dataset used for the experiment. Results are discussed in Section 4 and discussion in section 5. The conclusion and future work is discussed in Section 6.

## 3. Material & Methods

In this section we will discuss classifiers and dataset utilized for Road Severity Prediction. Figure 1 demonstrates the proposed methodology of data and work flow of this research work.

### 3.1. Modeling Methods

In this research, road accident severity analysis is performed by using ensemble learning models such as Voting Classifier (VC) based on Logistic Regression (LR) and Stochastic Gradient Descent (SGD), Random Forest (RF), AdaBoost Classifier (ADC), Extra Tree Classifier (ETC) and Gradient Boosting Machine (GBM). LR uses logit, that is natural logarithm to measure likelihood ratio of dependent variable as 1 in case of serious accident and 0 for the opposite case of minor accident. Probability of accident is represented by p and given by :

$$Y = logit = \ln \frac{P}{1-P} \not= \beta \ X \qquad (1)$$

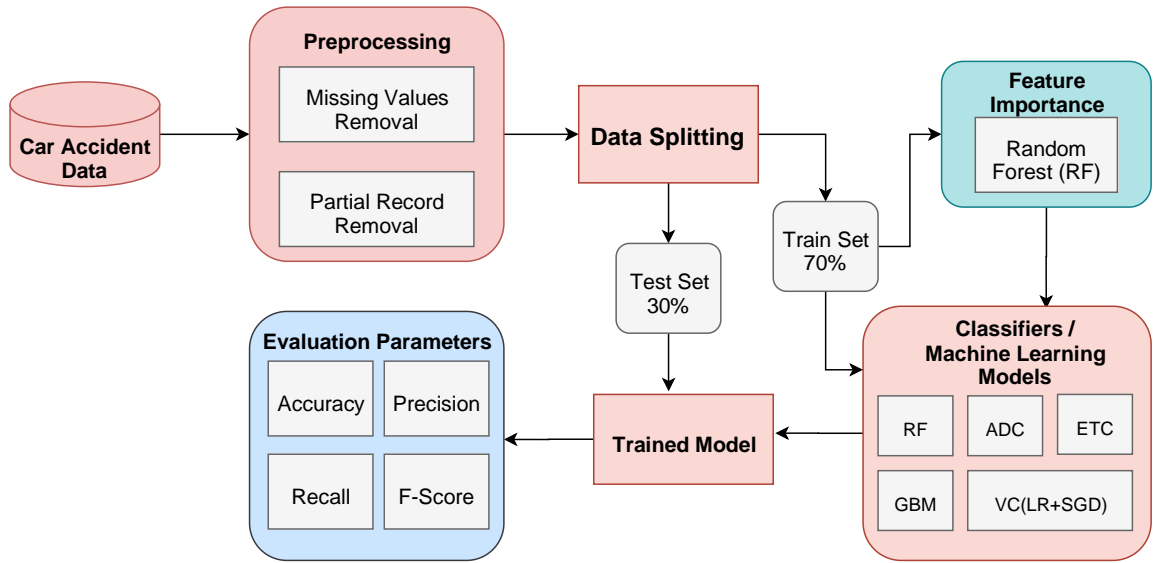

Figure 1    Proposed Methodology Diagram

Where Y is the variable measuring severity of accident. If accidental severity is serious y will be 1 and if severity is minor y will be 0. X represents independent variables and $\beta$ is to be estimated Tay, Rifaat, and Chin (2008). Stochastic gradient descent (SGD) is an iterative approach and optimizes objective function by selecting smoothness in terms of properties Bottou (2010). Actual gradient obtained from the dataset is replaced by an estimate which is calculated from the subset of the dataset. It achieves faster iteration by low convergence trade off and reduces the computational burden. It has been extensively used in machine learning and deep learning problems. Voting classifier is an ensemble combination of individual classifiers and combines prediction results of classifiers and could achieve better results than single classifiers Zhang et al. (2014a). This research utilizes the voting ensemble of LR and SGD to predict road accident severity.

In the past, many ensemble techniques have been proposed by researchers but most common are bagging Breiman (1996) and boosting Schapire (1999). Random forest (RF) was first developed by Breiman Breiman (2001). In RF algorithm if N number of trees are built by RF then four steps are involved in N iteration. Step 1 involves training of data using bootstrap dataset, bootstrap dataset is subset of original dataset. Step 2 involves tree generation and at step 3 attributes are selected randomly. Finally in step 4 final prediction is based on the tree result selected on the basis of majority voting Sekhar, Madhu et al. (2016). Working of RF is shown as follows.

$$p = mode\{T_1(y), T_2(y), ..., T_m(y)\} \qquad (2)$$

$$p = mode\left\{\sum_{m=1}^{m} T_m(y)\right\} \qquad (3)$$

Here p is the final prediction, calculate by majority votes of trees $T_1$, $T_2$ and $T_m$ Biau and Scornet (2016).

AdaBoost is a short form of adaptive boosting. It is also an ensemble model based on decision trees. AdaBoost Classifier (AC) is popular for being the first algorithm in adoption of weak learners Zhang et al. (2014b). AC trains weak learners recursively on duplication of original dataset where weak learners focus on outliers Freund and Schapire (1997). It is a meta classifier and trains weak learners on the same feature set but with different weights. AC outperformed in many classification tasks Sevilla-Noarbe and Etayo-Sotos (2015), Zitlau et al. (2016).

Extra Tree Classifier (ETC) also uses a random subset of features to split nodes of trees. But it builds trees using a complete sample unlike RF and randomly selects a cut point to split a node. ETC utilizes multi linear approximation instead of piecewise constant for RF. Extra randomization of ETC makes it superior in terms of performance than RF and base learners' mistakes are less correlated with each other. ETC showed better performance than RF in terms of accuracy in Geurts, Ernst, and Wehenkel (2006).

Gradient Boosting Machine (GBM) is based on boosting and a powerful ensemble model to perform classification. It uses ensemble of weak learners specifically decision trees for prediction Natekin and Knoll (2013), Friedman (2000). Weak learners are converted to strong one in boosting technique and every new tree is fit on modified form of trees. GBM uses gradient in loss function, which measures how efficiently the model coefficient fit the data.

In analysis of ensemble models' performance accident dataset is randomly divided into two groups, training and test dataset with the ratio of 70:30. Parameter tuning values are presented in the table. Significance of each variable represents the degree of its influence on accidental severity. In other words, more significant variables will affect accident severity for sure. Generally more frequently selected variable for split the higher score it will get for significance. Variable importance score can be calculated by RF by calculating prediction error, if variable values are permuted against out of bag instances. Significance score is measured for each tree, averaging the all ensemble and then by dividing it by standard deviation Shaikhina et al. (2019).

### 3.2. Performance evaluation parameters

The evaluation of models is an important task in classification, and different parameters have been represented in this regard. This study makes use of accuracy, precision, recall, and F-score, which are among the most commonly applied evaluation metrics. Formally, accuracy is used as the correctness of prediction, and is calculated as:

$$Accuracy = \frac{Number\ of\ correct\ predictions}{Total\ number\ of\ predictions} \quad (4)$$

whereas for binary classification, accuracy can also be calculated in terms of positives and negatives, as follows:

$$Accuracy = \frac{TP+TN}{TP+TN+FP+FN} \quad (5)$$

where *TP*, *TN*, *FP*, and *FN* represent true positive, true negative, false positive and false negative and are defined as follows Umer et al. (2020).

**True Positive (TP):** TP shows the positive predictions of a class that are correctly predicted by the classifier.

**True Negative (TN):** These are the negative predilections of a class that are correctly labeled by the classifier.

**False Positive (FP):** FP shows the negative predictions of a class that are incorrectly labeled as positive by the classifier.

**False Negative (FN):** These are positive predictions of a class that are incorrectly labeled as negative by the classifier.

Precision is referred to as the exactness of a classifier and tells what percentage of all tuples are labeled positive which are actually positive. It is calculated as:

$$Precision = \frac{TP}{TP+FP} \tag{6}$$

Recall on the other hand, is often called as the measure of completeness and it presents the percentage of true positive tuples which are labeled correctly. It is calculated as:

$$Recall = \frac{TP}{TP+FN} \tag{7}$$

F-score is a statistical analysis measure of classification, which considers both precision and recall
of the classifier and computes a score between 0 and 1 Umer et al. (2020b). It shows the effect of both precision and recall and is calculated as:

$$Fscore = 2\frac{precision.recall}{precision+recall} \tag{8}$$

### 3.3. Dataset

**3.3.1. Selection of Accident Dataset:** This is a countrywide car accident dataset, which covers 49 states of the USA. The accident data contains records from February 2016 to June 2020. There are about 3.5 million accident records in this dataset. The dataset contains 49 columns which are 'ID', 'Source', 'TMC', 'Severity', 'Start Time', 'End Time', 'Start Lat',

'Start Lng', 'End Lat', 'End Lng', 'Distance(mi)', 'Description', 'Number', 'Street', 'Side', 'City', 'County', 'State', 'Zipcode', 'Country', 'Timezone', 'Airport Code', 'Weather Timestamp', 'Temperature(F)', 'Wind Chill(F)', 'Humidity(%)', 'Pressure(in)', 'Visibility(mi)', 'Wind Direction',

'Wind Speed(mph)', 'Precipitation(in)', 'Weather Condition', 'Amenity', 'Bump', 'Crossing',

'Give Way', 'Junction', 'No Exit', 'Railway', 'Roundabout', 'Station', 'Stop', 'Traffic Calming', 'Traffic Signal', 'Turning Loop', 'Sunrise Sunset', 'Civil ₋Twilight', 'Nautical ₋Twilight', and 'Astronomical Twilight'. The number of records against each target class is shown in figure 2. The value '1' indicates the least severity while the value '4' tell us the accident is severe.

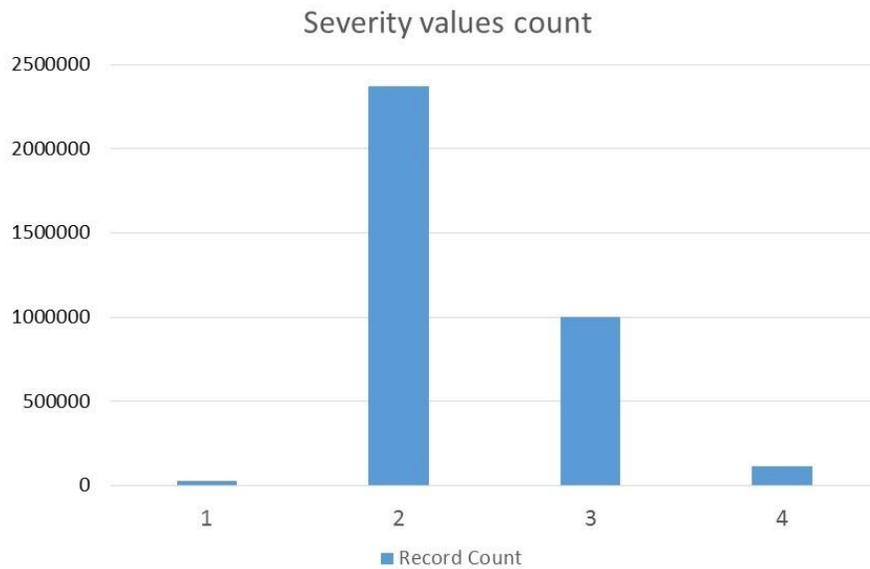

Figure 2     Countplot showing class-wise data distribution

**3.3.2. Data Visualization:** Data Visualization helps to understand the hidden patterns lying inside the dataset. It helps to get more details about the dataset by visualizing the characteristics of the attributes. Figure 3 shows the ratio of missing values of each column in the dataset. Figure 4 expresses the impact of top 5 weather conditions for accidents. Figure 4 concludes that most of the time when the accidents happens than the weather is clear.

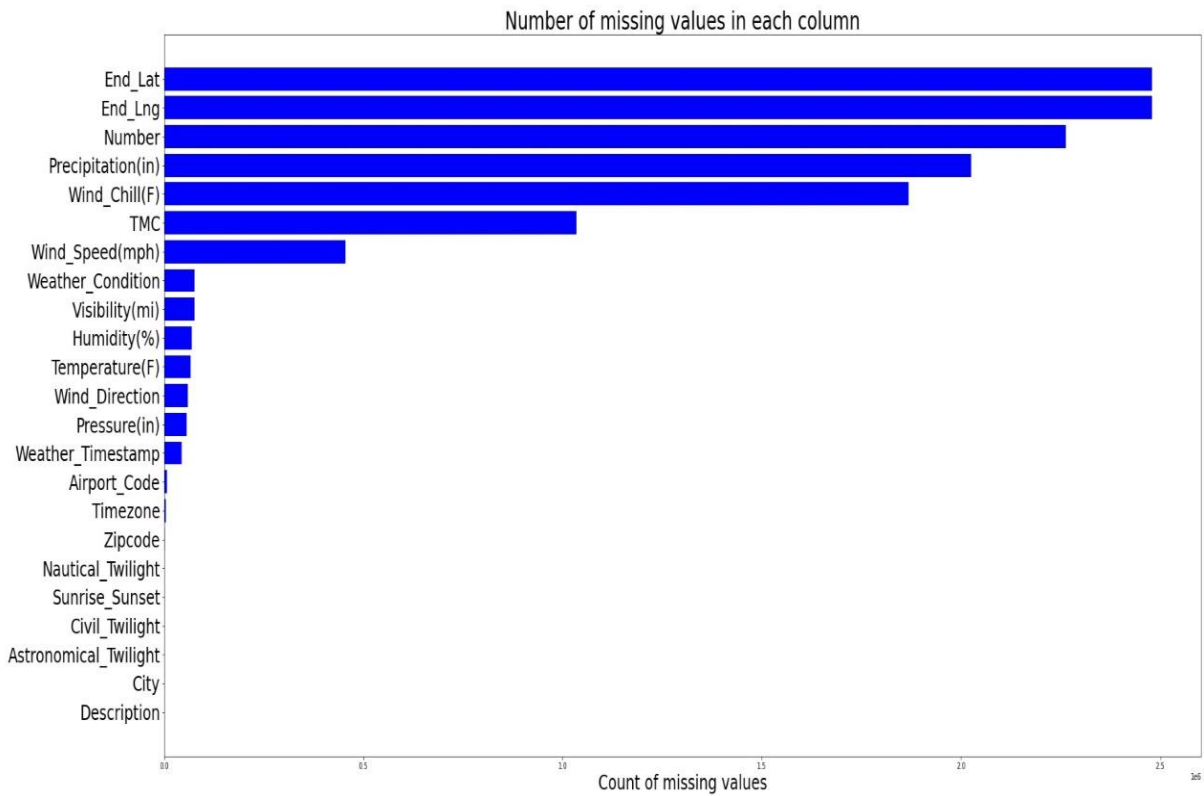

Figure 3        Countplot of missing values of each column.

**3.3.3. Data preprocessing:** Datasets contain unnecessary data in raw form that can be unstructured or semi-structured. That unnecessary raw data increases time of training the model but decreases performance of the model. Preprocessing plays a significant role in improving the performance of machine learning models and saving computational resources. Text preprocessing boost the prediction accuracy of the model Kalra and Aggarwal (2018). We performed following steps in preprocessing; missing values removal, partial records removal.

# 4.    Results

This section presents results after executing ensemble models Voting Classifier (LR+SGD), RF, AC, ETC and GBM on US road accident dataset. Significant variables identified by RF are illustrated in section 4.1. In this study from 48 original feature set variables, 20 important feature sets are identified by RF. Section 4.2 presents the comparison of accuracy by both sets of variables (original in section 4.2.1 and important in section 4.2.2) by comparing results of ensemble models.

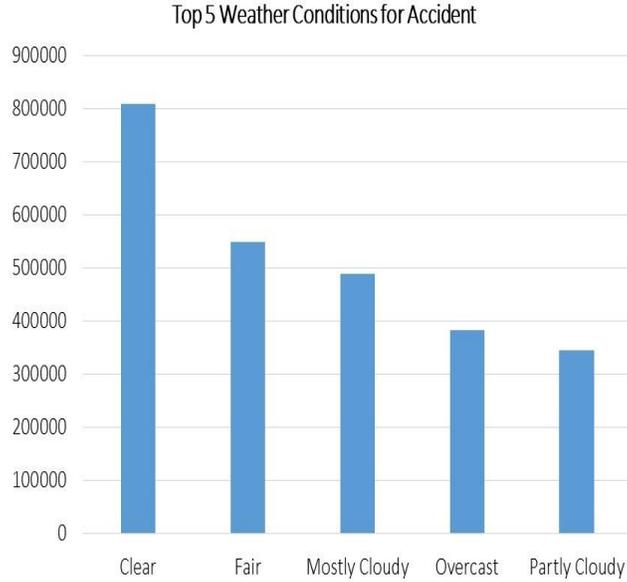

Figure 4   Top 5 weather conditions for accident.

## 4.1. Significant features

In this step, RF with the input of all 48 feature variables of the whole dataset identified top 20 significant features. Identified important features are presented in the figure 5.

## 4.2. Comprehensive Comparison of Predictive performance of Tree based and Statistical Ensemble Models

**4.2.1. Classification results with input of original 48 variables** Classification results of ensemble machine learning models (VC(LR+SGD), RF, AC,ETC and GBM) using all 48 features is presented in Table 1. It can be clearly observed that RF achieved 0.744 accuracy value, 0.784 precision and 0.790 recall, which are highest value among all other models. In addition ETC produce 0.728% accuracy, 0.698 precision which are highest values after RF. While VC(LR+SGD) achieved 0.789 recall which is second highest and 0.740 f-score value which is highest among all models.

Table 1   Classification result of all machine learning models using all features.

| Models | Accuracy | Precision | Recall | F-Score |
|---|---|---|---|---|
| Voting Classifier | 0.722 | 0.692 | 0.789 | 0.740 |
| Random Forest | 0.744 | 0.784 | 0.790 | 0.722 |
| AdaBoost Classifier | 0.704 | 0.682 | 0.711 | 0.696 |
| Extra Tree Classifier | 0.728 | 0.698 | 0.754 | 0.726 |
| Gradient Boosting Machine | 0.714 | 0.672 | 0.741 | 0.706 |

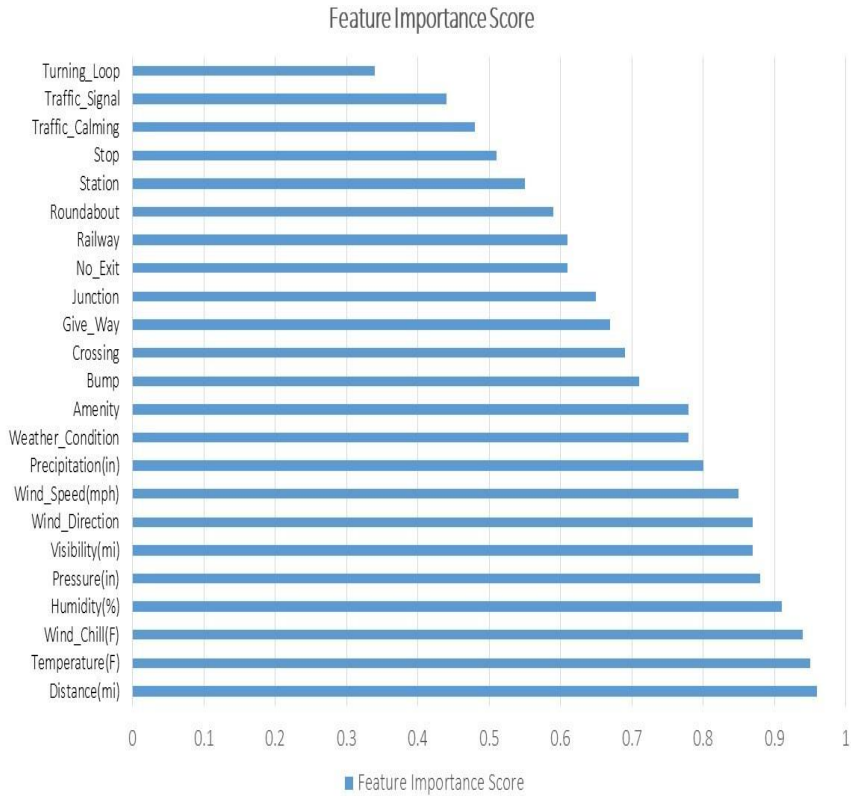

Figure 5   Feature importance calculated using random forest.

**4.2.2. Classification results with input of most significant 20 variables** Table 2 shows the accuracy, precision, recall and F-score of classification with significant features calculated using random forest features importance. It can be seen clearly that classification result of all models improved in this case.This means that utilizing significant features help is reducing extra noise from variables which increase the classification results as shown in table 2. On the other side, it also reduces extra cost of collection of accident data. Instead of collecting data of 48 features, there is need to collect only 20 features. In addition it can be observed clearly that Random Forest classifier achieved the highest accuracy value 0.974. RF achieved the highest precision value with 0.954. recall is 0.93, and 0.942 F-score.

Performance of voting classifier also improved after using significant features identified by RF. and it achieves 0.962 accuracy value which is second highest after RF. ETC achieved highest precision and highest f-score value with 0.928 and 0.916 respectively after RF. GBM achieved highest recall with 0.921 value using 20 significant features to predict accident severity.

Table 2    Classification result of all machine learning models using significant features.

| Models | Accuracy | Precision | Recall | F-Score |
|---|---|---|---|---|
| Voting Classifier | 0.962 | 0.912 | 0.919 | 0.915 |
| Random Forest | 0.974 | 0.954 | 0.930 | 0.942 |
| AdaBoost Classifier | 0.944 | 0.922 | 0.901 | 0.911 |
| Extra Tree Classifier | 0.917 | 0.928 | 0.904 | 0.916 |
| Gradient Boosting Machine | 0.921 | 0.902 | 0.921 | 0.911 |

## 5. Discussions

We conducted the comparison of Tree based ensemble models (Random Forest, AdaBoost, Extra Tree Classifier and Gradient Boosting Machine) and ensemble of regression algorithms (Voting classifier (LR+SGD)) to measure severity of road accidents. We also identified 20 significant features by Random forest which are almost half of the all available features of the dataset. In our experiment, we used all available features of the dataset as input for the all ensemble models in the first phase. While in the second phase of experiment, we used most significant features identified by RF as input for all ensemble models.

Empirical results for accidental severity prediction are summarized in four main findings. First in terms of accuracy, RF outperformed among all above mentioned ensemble learning models to predict accident severity. Accuracy of RF (0.974), using 20 significant features as input, is the highest in this study. Accuracy results of all ensemble models using all available features and significant features are presented in Figure 6. Using all features as input in Voting Classifier (LR+SGD) achieved 0.722 accuracy value, while by using significant features it also improved with 0.962 accuracy value. It can be noticed thay by utilizing significant features identified by tree based model RF performance of ensemble of regression models has also improved. It can be observed clearly from figure 6 that accuracies of all ensemble models are improved more than 20% using significant features as input rather than using all features as input.

Second in terms of precision, RF achieved better result with significant difference as compared to voting classifier as shown in figure 7. RF achieved higher values with 0.784 precision value using all features as well as with 0.954 precision value using significant features identified by RF. The Voting classifier achieved 0.692 precision value using all features as input and 0.912 precision value using significant features which are less than RF precision score. Tree ensemble models RF, AC, ETC are achieving high precision scores (0.954, 0.922 and 0.928 respectively) using significant features as compared to voting classifier (0.912). But precision score of GBM is lower than voting classifier using all features (0.672) and using significant features (0.902).

Third in terms of recall (sensitivity), RF achieves 0.93 recall score using significant features, which is the highest recall value for predicting accident severity. AC and ETC achieved almost similar recall values using significant features which are 0.901 and 0.904 respectively. GBM achieved less recall score with 0.741 than voting classifier with 0.789 using all features. But GBM also achieved higher recall value with 0.921 than voting classifier with 0.919 score using significant features. When ensemble models are input with significant variables identified by RF improved their recall score. Recall score is presented in figure 8.

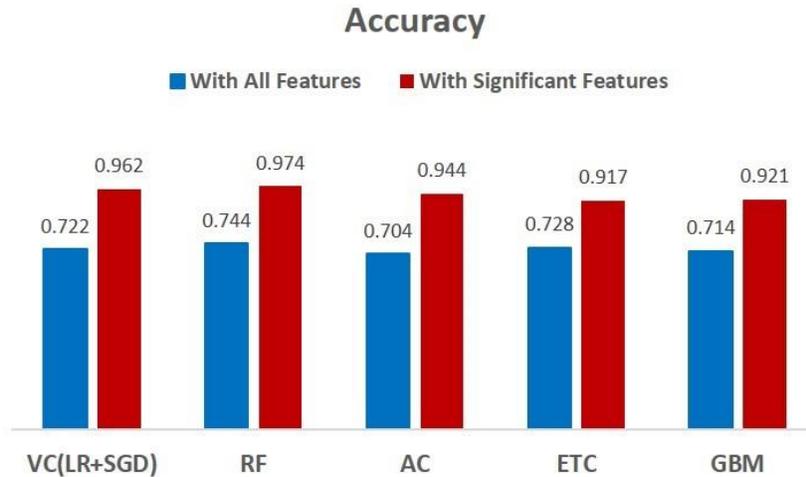

Figure 6    Accuracy of all Ensemble learning models.

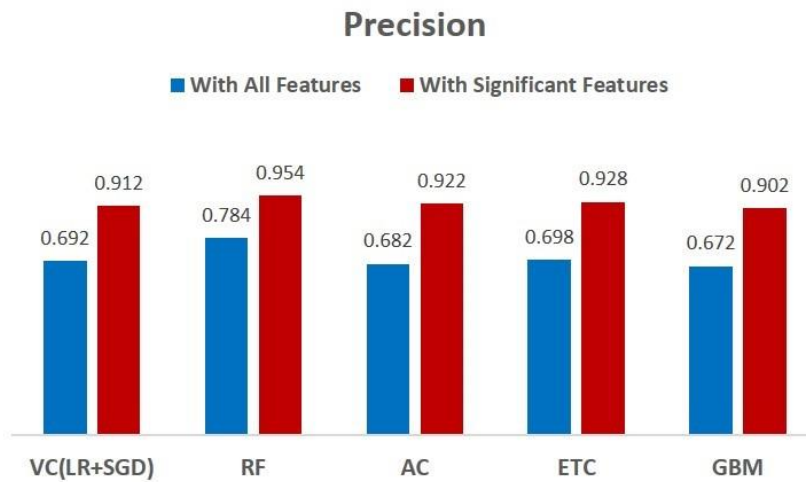

Figure 7    Precision of all Ensemble learning models.

Fourth in terms of F-score, which is another important evaluation measure and provides balance between precision and recall scores, RF outperformed other aforementioned ensemble models. From figure 9, it can be seen that F-score or RF (0.722) is lower than Voting classifier (0.74) using all features. But by using significant features as input RF achieved the highest F-score with 0.942 value.

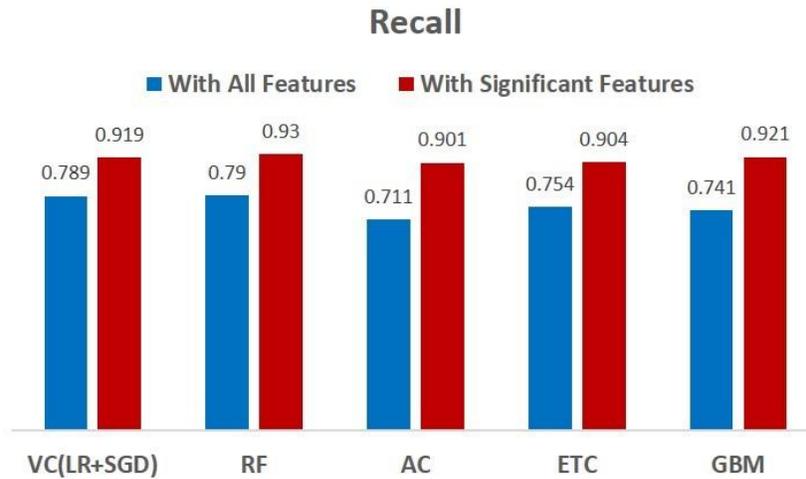

Figure 8        Recall of all Ensemble learning models.

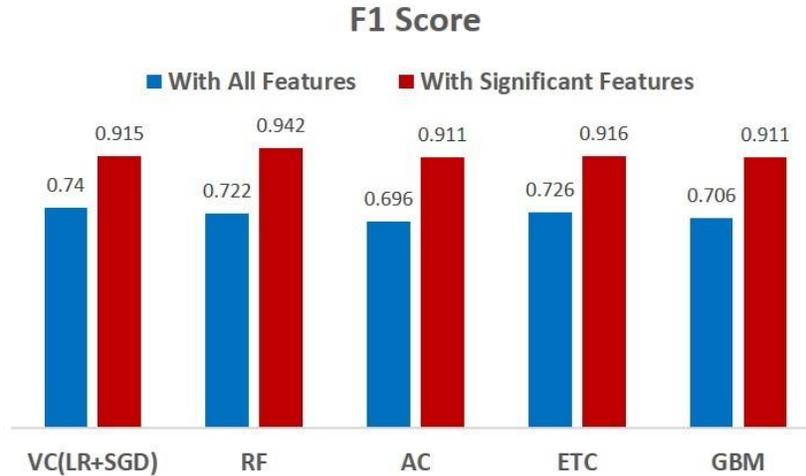

Figure 9        F-Score of all Ensemble learning models.

In general, based on the summary of the results, if the primary focus is the overall performance of the models in predicting road accident severity, RF achieved highest results using significant features. Significant features, that are a subset of the original feature set, are identified by RF. By using significant features as input by ensemble models not only significantly improve accuracy but also improved precision , recall and f-score. RF in combination with significant features not only improving prediction performance but also reduce the cost of the data collection. There are 48 features regarding road accident dataset to measure severity of the accident. By considering 20 important variables identified by RF significantly improve the prediction process of accident severity. Performance of ensemble models is also compared using all features and using important features as input. Tree based ensembles showed better performance to predict accidental severity due to their ability to learn non linear solutions and these models scale well on large datasets. Management should pay more attention to the 20 most important features affecting accident severity.

## 6. Conclusion

In this paper, experimental results explained that classification results of RF are higher than AC, ETC, GBM and voting classifier (LR+SGD). Most significant features identified by RF are also used as input to the ensemble models and also promote accuracy, precision, recall and f-score of all ensemble models but RF again outperformed with significant difference. Therefore it can be said that RF is the most efficient and effective model among all ensemble models and showed consistent results in predicting accident severity.

On the other hand, identification of significant features from overall features was focused to measure their correlation with the road accidents. Influence of significant variables on prediction performance results of ensemble models is also evaluated in this study. In the first phase experiment is performed by using all features, while in the second phase experiment is performed by using significant variables identified by RF. Important variables improved accuracy, precision, recall and f-score of all ensemble models. Therefore it can be said that important features identified by RF can help in boosting prediction performance of the models and also reduce the cost of data collection. The result shows that distance between vehicles is the most important feature affecting severity of the accident so road authorities can take preventive measures by focusing on important features identified by RF.

**Acknowledgments**

This work is supported by Department of Computer Science, Khwaja Fareed University of Engineering and Information Technology, Rahim Yar Khan, Pakistan.